\documentclass{ifacconf}

\usepackage{graphicx}      
\usepackage{natbib}        

\usepackage{glossaries}
\usepackage{amsfonts} 
\usepackage{amssymb} 
\usepackage{amsmath}
\usepackage{mathrsfs}
\usepackage{multicol}
\usepackage{graphicx}
\usepackage{booktabs}
\usepackage{tcolorbox}
\usepackage{tabularx}
\usepackage{array}
\usepackage{colortbl}
\usepackage[utf8]{inputenc}
\usepackage{xcolor}
\usepackage{color}
\usepackage{setspace}
\usepackage{stackengine}
\usepackage{subfigure}
\usepackage{algorithm}
\usepackage{algpseudocode}
\usepackage{bm}
\usepackage{tikz}
\usetikzlibrary{shapes,arrows}
\usepackage{verbatim}

\usepackage{lipsum}

\usetikzlibrary{positioning}

\tikzstyle{block} = [draw, fill=pink, rectangle, 
minimum height=3em, minimum width=6em,node distance=2cm]
\tikzstyle{sum} = [draw, fill=pink, circle, node distance=1cm]
\tikzstyle{input} = [coordinate]
\tikzstyle{output} = [coordinate]
\tikzstyle{pinstyle} = [pin edge={to-,thin,black}]

\newcommand{\stl}{\textbf{STL}}
\newtheorem{assumption}{\textbf{Assumption}}
\newtheorem{problem}{\textbf{Problem}}
\newtheorem{remark}{\textbf{Remark}}
\newtheorem{definition}{\textbf{Definition}}
\newtheorem{theorem}{\textbf{Theorem}}
\newtheorem{lemma}{\textbf{Lemma}}

\newcommand{\always}{\Box}
\newcommand{\eventually}{\Diamond}
\newcommand{\until}{\mathbin{\sf U}}

\newcommand\Ali[1]{{\color{black}#1}}

\newcommand\barbelow[1]{\stackunder[1.2pt]{$#1$}{\rule{.8ex}{.075ex}}}
\begin{document}
\begin{frontmatter}

\title{Data-Driven Verification under Signal Temporal Logic Constraints\thanksref{footnoteinfo}} 

\thanks[footnoteinfo]{This work was supported in part by the H2020 ERC Starting Grant AutoCPS (grant agreement No. 804639).}

\author[First]{Ali Salamati} 
\author[Second]{Sadegh Soudjani} 
\author[First,Third]{Majid Zamani}

\address[First]{Computer Science Department, Ludwig Maximilian University of Munich, Germany (e-mail: ali.salamati@lmu.de)}
\address[Second]{School of Computing,  Newcastle University, Newcastle upon Tyne, United Kingdom (e-mail: sadegh.soudjani@newcastle.ac.uk)}
\address[Third]{Computer Science Department, University of Colorado Boulder, USA (e-mail: majid.zamani@colorado.edu)}

\begin{abstract}                
We consider systems under uncertainty whose dynamics are partially unknown. Our aim is to study satisfaction of temporal logic properties by trajectories of such systems. We express these properties as signal temporal logic formulas and check if the probability of satisfying the property is at least a given threshold. Since the dynamics are parameterized and partially unknown, we collect data from the system and employ Bayesian inference techniques to associate a confidence value to the satisfaction of the property. The main novelty of our approach is to combine both data-driven and model-based techniques in order to have a two-layer probabilistic reasoning over the behavior of the system: one layer is related to the stochastic noise inside the system and the next layer is related to the noisy data collected from the system. We provide approximate algorithms for computing the confidence for linear dynamical systems.    
\end{abstract}


\end{frontmatter}
\section{Introduction}
Formal methods have been vastly used in computer science to provide correctness guarantees on the expected behavior of a program. Most of these formal techniques have been developed for finite-state models \citep{beyer2018unifying,beyer2011cpachecker}. In order to fully utilize the advantages of formal techniques in real physical applications, one needs to first construct a sufficiently precise model of the system.
Usually, it is hard to model a system accurately. Besides, the dynamics of a system may vary throughout the course of time. In such cases, statistical model checking can be beneficial if all the states of the system can be measured \citep{sen2004statistical,clarke2011statistical,sen2005statistical}. However, statistical model checking usually needs a large number of experiments, is not able to deal efficiently with uncertainties in the system, and is not able to handle synthesis problems directly \citep{sen2005statistical}.

A data-driven approach was developed in \cite{sadraddini2018formal} for control of piecewise affine systems with additive disturbances against signal temporal logic ($\stl$) properties. The work in \cite{bartocci2014data} exploits concepts from formal modeling and machine learning to develop methodologies that can identify temporal logic formulae that discriminate different stochastic processes based on observations. The results in \cite{chou2019bayesian} propose an approach to approximate the posterior distribution of unknown parameters for a nonlinear and deterministic system.

$\stl$ properties are introduced and used in the literature including the works by \cite{raman2015reactive} and \cite{fainekos2006robustness}. The work in \cite{sadigh2016safe} introduces a new definition for the probabilistic $\stl$ that assigns probabilities to the atomic propositions and then combines them through Boolean operators. The results in \cite{farahani2018formal} utilize probabilistic $\stl$ properties to design a control strategy for Barcelona wastewater system. Satisfaction of properties expressed in linear temporal logic on finite traces for linear time-invariant (LTI) systems is investigated in \cite{haesaert2015data,haesaerta2016data} by using Bayesian inference. The proposed approach in \cite{polgreen2017automated} applies
Bayesian inference to parametric Markov chain.

Recently, researchers studies data-driven techniques for formal policy synthesis of dynamical systems due to their applicability to high dimensional spaces. A data-driven approach is proposed in \cite{shmarov2019automated} for synthesis of safe digital controllers for sampled-data stochastic nonlinear systems. The results in \cite{lavaei2020formal} use model-free reinforcement learning for policy synthesis of dynamical systems with finite-horizon properties under continuity assumptions on the dynamics of the system.
Finally, \cite{kazemi2020formal} uses reinforcement learning for satisfying all (infinite-horizon) linear temporal logic properties with convergence guarantees and without any continuity assumption on the dynamics of the system. 

In this work, a Bayesian framework is introduced in order to give a probabilistic confidence measure over an $\stl$ property for a set of parameterized models of stochastic systems. 
In our approach, a prior knowledge of the system accompanied by the collected data from the system are leveraged together to improve the confidence of satisfaction for the properties of interest expressed as $\stl$ formulas. Our main objective is to combine both data-driven and model-based techniques for stochastic systems in order to verify the system against probabilistic $\stl$ properties.
The results are demonstrated for partially unknown linearly parameterized models of stochastic systems.

Our approach considers a probability threshold as lower bound for the satisfaction of $\stl$ property by the stochastic trajectories of the system. We under-approximate the feasible parameter set of the probabilistic constraint by transforming them into algebraic inequalities.
Then, a confidence value is computed using the obtained feasible set and the distribution of the parameter is updated based on collected data from the system. We also propose relaxation of algebraic inequalities in order to reduce the conservativeness of the under-approximation.

In summary, our contributions are threefold. First, we provide a probabilistic confidence of satisfying an $\stl$ property for a stochastic LTI system. Second, we provide an approach to under-approximate the feasible domain of the $\stl$ chance constraint that is less conservative. Third, two numerical approaches are developed in order to compute the feasible set of parameters.

The structure of the paper is as follows. In Section~\ref{sec:prbstate}, definitions and assumptions are introduced regarding the stochastic confidence value and techniques on collecting data. In Section~\ref{sec:Bayes}, Bayesian inference is developed for systems affected by both measurement and process noises. Section~\ref{sec:STL} demonstrates a technique in order to under-approximate the feasible domain of the probabilistic $\stl$ constraints. Section~\ref{sec:verification} shows how to compute the feasible set of parameters for stochastic LTI systems. Finally, illustration of the approach on a case study is presented in Section ~\ref{sec:casestudy}. We did not include the proofs due to space limitations and will be included in an online arXiv version of the paper.

\section{preliminaries and problem formulation}
\label{sec:prbstate}
In this section, we make a clear overview of our problem and the proposed approach to tackle that. Consider the original system $\mathbf{S}$ in Fig.~ \ref{fig:block_diag1} which is affected by both process noise $w(t)$ and measurement noise $e(t)$. Assume that there are parametric models $M(\theta)$ of the original system in which $\theta$ comes from a parameter set $\Theta$. This set of models is described as $\Omega=\{M(\theta)\mid\theta\in\Theta\}$. 
\begin{assumption}
    It is assumed that there is a true parameter $\theta_{\textit{true}}$ such that $M(\theta_{\textit{true}})$ describes the behavior of the original system $\mathbf{S}$. This true parameter is unknown in general.
\end{assumption}

Consider a property $\psi$ defined over trajectories of the system $\mathbf{S}$. We assume this property belongs to the class of $\stl$ properties which will be defined in Subsection~\ref{subsec:STL}. We denote satisfaction of $\psi$ by the trajectories of the system with $\mathbf{S}\models\psi$.  
We intend to give a confidence value for the satisfaction of a probabilistic $\stl$ property $\psi$ for a system $\mathbf{S}$ by combining Bayesian inference and model-based techniques. We consider both process noise over the dynamics of the system, and measurement noises over outputs of the system. Hence, we can consider the overall system as a two-layer noise framework, which is illustrated in Fig. ~\ref{fig:block_diag1}. By collecting and analyzing data, we plan to provide a confidence value for the satisfaction of the desired property.
\subsection{Data Collection}
As it is seen from Fig. ~\ref{fig:block_diag1}, we have a stochastic system $\mathbf{S}$ whose output is mixed with a measurement noise $e(t)$. Let us define the set of data collected from the system $\mathcal{D}=\{\tilde{u}_{exp}(t),\tilde{y}_{exp}(t)\}_{t=0}^{\textbf{N}_{\textbf{exp}}-1}$, in which $\tilde{u}_{exp}(t)$ and $\tilde{y}_{exp}(t)$ are input-output pairs for $\textbf{N}_{\textbf{exp}}$ measurements. In general, it is assumed that we can excite the system with any desirable input signal but within the acceptable range of inputs.
\begin{remark}
The acceptable range of input for collecting data can be allowed to be larger than the range of inputs for the verification of the $\stl$ property in order to provide a more precise confidence.  
\end{remark}
\tikzstyle{block} = [draw, fill=pink, rectangle, 
minimum height=3em, minimum width=6em]
\begin{figure}
    \centering
    \begin{tikzpicture}[auto, node distance=1.5 cm,>=latex',scale=.1]
    \node [input, name=input] {};
    
    \node [block, right of=input, pin={[pinstyle]above:$w(t)$},
    node distance=2cm] (system) {$\mathbf{S}$};
    \node [sum, right of=system, pin={[pinstyle]above:$e(t)$},
    node distance=2cm] (sum) {};
   \node [output, right of=sum] (output1) {};
    \node [output, right of=output1] (output2) {};
    \node [block, below of=system] (measurements) {Data Analyzer};
    
    \draw [draw,->] (input) -- node [pos=-.1] {$ u(t)$}(system);
    \draw [->] (output1) -- node {$y(t)$} (output2);
    \draw [-] (sum) -- (output1);
    \draw [->] (system) -- node [name=y] {$\hat{y}(t)$}(sum);
    \draw [->] (output1) |- (measurements);
    \draw [->] (input) |- node[pos=0.99] {} 
    node [near end] {} (measurements);
    \end{tikzpicture}
    \caption{Two-layer noise framework and data collection setup} 
    \label{fig:block_diag1}
\end{figure}
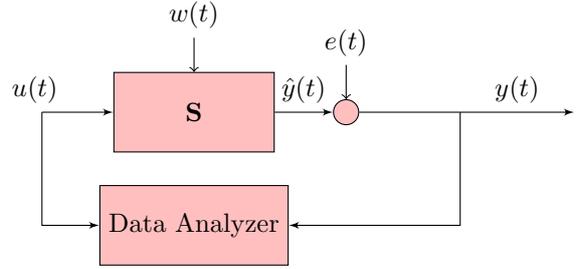

\begin{assumption}
    Both process and measurement noises are considered independent and identically distributed. Besides, they are not correlated to the input signals. Initial state vector $x(0)$ is considered to be known.
\end{assumption}
\subsection{ Stochastic Bayesian Confidence}
Satisfaction of a property $\psi$ for a deterministic system can be considered as a binary value over the parameter space $\Theta$. If we have $\Omega$ as the set of parameterized deterministic models over the whole parameter space $\Theta$, satisfaction function for the deterministic system can be defined as $g_{\psi} : \Theta\to\{0,1\}$ in which \Ali{$g_{\psi}(\theta)\equiv M(\theta)\models\psi$ }. This confidence value can be only zero or one.
 If the system is affected by the process noise, the satisfaction of the desired property can be explained by a probabilistic measure. Now, we can define a threshold on the probability of satisfaction of a property $\psi$ as
 \begin{equation}
 \label{eq:chance_const}
 \textbf{Pr}(M(\theta)\models\psi)\geq 1-\delta,
 \end{equation}
 where $\delta\in(0,1)$. Now we can assign a satisfaction function $f_{\psi}^\delta$ to the above chance constraint which is again a binary function on the parameter space $\Theta$.
\begin{definition}
            Consider $\Omega$ as the set of stochastic models $\textbf{M}(\theta)$ in which $\theta\in\Theta$, and let $\psi$ be a temporal logic formula (e.g. $\stl$). The stochastic satisfaction function $f_{\psi}^\delta : \Theta\to\{0,1\}$ is defined as:
    \begin{equation}
        f_{\psi}^\delta(\theta)=
        \begin{cases} 
        1 & \text{if $\textbf{Pr}(M(\theta)\models\psi)\geq 1-\delta$}, \\
        0 & \text{otherwise}.
        \end{cases}
        \label{eq:fsfc}
    \end{equation}
\end{definition}
Let $\textbf{Pr}(.)$ and $p(.)$ denote a probability value and a probability density function, respectively. Then, we can define a stochastic notion of confidence using Bayesian probability inference. It can be expressed as a distribution over the whole set of models $\Omega$.
\begin{definition}
    Given a property $\psi$ and a set of data $\mathcal{D}$, the notion of confidence for the stochastic system can be computed as:
    \begin{equation}
        \textbf{Pr}(\mathbf{S}\models\psi\mid\mathcal{D})=\int_{\Theta} f_{\psi}^\delta(\theta)\ p(\theta\mid\mathcal{D})d\theta,
        \label{eq:confidence}
    \end{equation}
where $ p(\theta\mid\mathcal{D})$ is a posteriori uncertainty distribution, given input-output pairs of data, and $f_{\psi}^\delta(\theta)$ is the stochastic satisfaction function defined in \eqref{eq:fsfc}.
\end{definition}
\subsection{Parametric LTI Systems}
\Ali{Note that the integral in \eqref{eq:confidence} is difficult to be tackled analytically in general. Therefore, we provide a computational approach suitable for \emph{linear time-invariant (LTI) systems} defined next.}

The nominal model for the stochastic system $\mathbf{S}$ is defined as:
\begin{equation}
\textbf{M}(\theta)\in \left\{\begin{array}{l}
x(t+1)=Ax(t)+B{u}(t)+Gw(t)\\ 
~~~\hat{y}(t,\theta)=C(\theta)x(t),
\end{array}\right.
\label{eq:parametricmodel}
\end{equation} 
where, $x(t)\in{\mathbb{R}^{n}}$, $y(t)\in{\mathbb{R}^{p}}$, and ${u}(t)\in\mathcal{U}\subset{\mathbb{R}^{m}}$, respectively. $\mathcal{U}$ is the set of valid inputs and is assumed to be bounded. We assume that matrices $A$ and $B$ are known. Signal $w(t)$ is the process noise with a zero-mean Gaussian distribution, which has a covariance matrix $\pmb{\Sigma_w}$. Due to collecting data, the stochastic system in Fig. ~\ref{fig:block_diag1} is also affected by the measurement noise as
\begin{equation}
\label{eq:full_dynamics}
{y}(t,\theta)=C(\theta)x(t)+e(t),
\end{equation}
 in which $e(t)\in{\mathbb{R}^{p}}$ is the measurement noise with a zero-mean Gaussian distribution, which has a covariance matrix $\pmb{\Sigma_e}$. Both process and measurement noises are assumed to be uncorrelated from the input signals. 
\subsection{Problem Statement}

Let us consider the framework in Fig. ~\ref{fig:block_diag1}, where the nominal system is affected by the process noise and its output is also affected by the measurement noise while collecting data. There is a set of parameterized models $\textbf{M}(\theta)$ for the stochastic system without the measurement noise. Also, we assume that there is a $\theta_{\textit{true}}$ from the parameter space $\Theta$ such that $M(\theta_{\textit{true}})$ describes the behaviors of the stochastic system.

\begin{figure}[h]
    \centering
    \begin{tikzpicture}[auto, node distance=1.5cm,>=latex',scale=.2]
    \node [input] (input1) {};
    \node [input, below of=input1] (input2) {};
    \node [input, below of=input2] (input3) {};
    \node [input, below of=input3] (input4) {};
    \node [block,above right =.25cm and .25cm of input2] (modelchecker) {Model Checker};
    \node [block,below right=.25cm and .25cm of input3] (dataanalyzer) {Data Analyzer};
    \node [block,below right =.5cm and .5cm of modelchecker] (confidencecomputation) {Confidence Computation};
    \node [output, right of=confidencecomputation]  (output) {};
    \draw [->] (input1) -| node[name=u1,above,pos=.1] {$\Theta$}   (modelchecker);
    \draw [->] (input2) -| node[name=u2,below,pos=.1] {$\psi$}    (modelchecker);
    \draw [->] (input3) -| node[name=u3,above,pos=.1] {$\tilde{u}_{exp}(t)$}  (dataanalyzer);
    \draw [->] (input4) -|  node[name=u4,below,pos=.1] {$\tilde{y}_{exp}(t)$}  (dataanalyzer);
    \draw [->] (modelchecker) -| node[name=u5,above,pos=.3] {$\Theta_\psi$}   (confidencecomputation);
    \draw [->] (dataanalyzer) -| node[name=u6,below,pos=.3] {$\textbf{p}(\theta\mid\mathcal{D})$}  (confidencecomputation);
    \end{tikzpicture}
    \caption{An overview of our proposed approach} 
    \label{fig:block_diag2}
\end{figure}
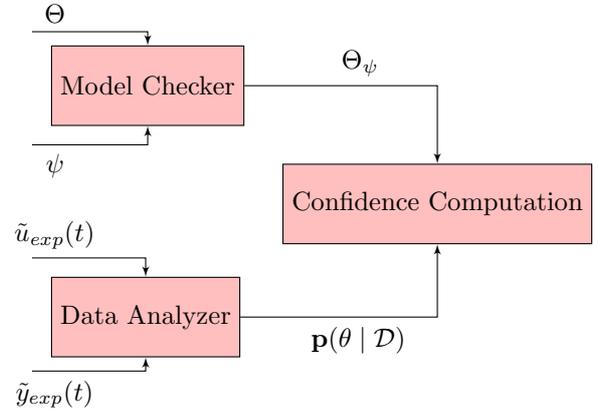

Assume that we have a prior knowledge of parameterized models for this system. This prior knowledge can be used in order to improve the posterior distribution function over the parameter space after collecting data from the system.
\begin{problem}\label{prob:prob1}
      \Ali{Given a parameterized LTI system in \eqref{eq:parametricmodel} together with the noisy output data in \eqref{eq:full_dynamics}, data set $\mathcal{D}$, and an $\stl$ property $\psi$, we aim at computing the confidence value in \eqref{eq:confidence}, with which the $\stl$ specification $\psi$ is satisfied independently of the input value.}
\end{problem}
This approach is depicted in Fig.~ \ref{fig:block_diag2}. In this figure, $\Theta$ is the whole parameter space. We denote by $\Theta_\psi$ the initial feasible set of parameters which their related parametric models satisfy the given probabilistic $\stl$ formula $\psi$. In addition, $p(\theta\mid\mathcal{D})$ denotes a posterior distribution function which is improved based on the collected data from the system, i.e., $\mathcal{D}=\{\tilde{u}_{exp}(t),\tilde{y}_{exp}(t)\}_{t=0}^{\textbf{N}_{\textbf{exp}}-1}$. The updated posterior distribution function will be leveraged in order to compute the confidence value using \eqref{eq:confidence}. Moreover, the prior information regarding appropriate parameters $\theta$, can be incorporated in order to achieve a more precise confidence.

\section{Bayesian inference}
\label{sec:Bayes}
We use Bayesian inference in order to provide the confidence of property satisfaction for parametric LTI systems. In many practical situations, we have an initial insight over the behaviors of the system that can be leveraged in order to increase our perception about the system. Bayesian inference is a powerful framework in order to incorporate this prior knowledge. Furthermore, the Bayesian framework is an efficient data-driven method. As it was mentioned before, confidence can be computed using (\ref{eq:confidence}). In (\ref{eq:confidence}), given the set of input-output data pairs, a posterior uncertainty distribution $p(\theta\mid\mathcal{D})$ can be inferred for the parameter $\theta$ by 
\begin{equation}
p(\theta\mid\mathcal{D})=\frac{p(\mathcal{D}\mid\theta)\;p(\theta)}{\int_{\Theta}p(\mathcal{D}\mid\theta)\;p(\theta)d\theta},
\label{eq:posterior}
\end{equation}
where $p(\theta)$ indicates a prior distribution over the whole parameter set $\Theta$ and comes from our initial knowledge of the system. Here, $p(\mathcal{D}\mid\theta)$ is the \emph{likelihood distribution function} which is computed based on our observations within the noisy environment.
Let us consider the set of data $\mathcal{D}=\{\tilde{u}_{exp}(t),\tilde{y}_{exp}(t)\}_{t=0}^{\textbf{N}_{\textbf{exp}}-1}$ in which $\tilde{u}_{exp}(t)$ and $\tilde{y}(t)_{exp}$ are input-output pairs for $\textbf{N} _{\textbf{exp}}$ measurements. The system gets excited with inputs $\tilde{u}_{exp}(t)$, and $\tilde{y}_{exp}(t)$ are the corresponding observed outputs of the system at time $t$ which are noisy. If the system is only affected by the measurement noise, observations can be assumed to be independent and identically distributed. In this case, the likelihood distribution $p(\mathcal{D}\mid\theta)$ can be computed simply as
$p(\mathcal{D}\mid\theta)=\prod_{t=0}^{\textbf{N} _{\textbf{exp}}-1}\;\; p(\tilde{y}_{exp}(t)\mid\theta).$
By considering the process noise, one can clearly observe that measurements will not be independent anymore.
In this scenario, we consider the likelihood distribution as a joint distribution function of all $\textbf{N} _{\textbf{exp}}$ measurements in the form of:
 \begin{equation}
 p(\tilde{y}_{exp}(0),\tilde{y}_{exp}(1),\ldots,\tilde{y}_{exp}(\textbf{N} _{\textbf{exp}}-1)\mid\theta),
 \end{equation}
where distributions for both measurement and process noises are assumed to be Gaussian with zero means and their corresponding covariances. We can consider this joint probability distribution function as a multi-variate Gaussian distribution function. The next theorem provides covariance matrix for the noisy outputs of the system.
\begin{theorem}
Consider the LTI model \eqref{eq:parametricmodel}-\eqref{eq:full_dynamics}. The joint distribution $p(\mathcal{D}\mid\theta)$ is multi-variate Gaussian with mean
\begin{align}
\label{eq:mean}
\bar{\boldsymbol{\mathbf{y}}}(\theta) = [\bar{y}(0);\cdots;\bar{y}(\mathrm{N} _{exp})],
\end{align}
and covariance matrix $\Sigma_{\tilde{\boldsymbol{\mathbf{y}}}}(\theta)$, 
where
\begin{align*}
\bar{y}(t)&:=C(\theta)A^tx(0)
+\sum_{i=0}^{t-1}C(\theta)A^{i}Bu(t-i-1)\\
\Sigma_{\tilde{\boldsymbol{\mathbf{y}}}}(\theta)& :=\mathbf{M(\theta)}\;\Sigma_{W}\;\mathbf{M(\theta)}^{T}+\Sigma_{E}.
\end{align*}
Matrices $\Sigma_{W}:=diag(\Sigma_w,\Sigma_w,\ldots,\Sigma_w)$ and $\Sigma_{E}:=diag(\Sigma_e,\ldots,\Sigma_e)$ are block diagonal.\\
$\mathbf{M}(\theta) \in \mathbb{R}^{(m\mathrm{N} _{exp}+m)\times(n\mathrm{N} _{exp})}$ is computable using matrices of the system as
\begin{equation*}
\resizebox{\linewidth}{!}{$
	\mathbf{M}(\theta)\!\!=\!\! \begin{bmatrix}
	0 & 0 & 0 & \cdots & 0 \\
	C(\theta)G & 0 &0 & \cdots & 0 \\
	C(\theta)AG     & C(\theta)G & 0 & \cdots & 0 \\
	C(\theta){A}^{2}G & C(\theta)AG & C(\theta)G&  \cdots & 0 \\
	\vdots  & \vdots&  \vdots      & \vdots &  \vdots\\
	C(\theta){A}^{\mathrm{N} _{exp}-1}G  & C(\theta){A}^{\mathrm{N} _{exp}-2}G  &  \cdots     & \cdots  &  C(\theta)G \\ 
	\end{bmatrix}.
	\label{eq:cov}
	$}
\end{equation*}
\label{theo:covarince}
\end{theorem}
The previous theorem results in a symmetric parametric covariance matrix, $\boldsymbol{\Sigma_{\tilde{\boldsymbol{\mathbf{y}}}}(\theta)}$, for $\textbf{N} _{\textbf{exp}}$ measurements of the system. Now, the joint Gaussian distribution function for $\textbf{N} _{\textbf{exp}}$ measurements is given by: 
\begin{equation}
\resizebox{\linewidth}{!}{$
\begin{split}
   &p(\tilde{y}_{exp}(0),\tilde{y}_{exp}(1),\ldots,\tilde{y}_{exp}(\textbf{N} _{\textbf{exp}}-1)\mid\theta) = \\
  &\frac{1}{|\boldsymbol{\Sigma_{\tilde{\boldsymbol{\mathbf{y}}}}(\theta)}|^{\frac{1}{2}} (2 \pi)^{\frac{\textbf{N} _{\textbf{exp}}}{2}}} \exp\left\{ -\frac{1}{2} (\tilde{\boldsymbol{\mathbf{y}}}-\hat{\boldsymbol{\mathbf{y}}}(\theta))^\textbf{T}\;\boldsymbol{\Sigma_{\tilde{\boldsymbol{\mathbf{y}}}}(\theta)}^{-1}(\tilde{\boldsymbol{\mathbf{y}}}-\hat{\boldsymbol{\mathbf{y}}}(\theta))\right\},
\label{eq:multivariate}
\end{split}
$}
\end{equation}
where, $\tilde{\boldsymbol{\mathbf{y}}}$ and $\hat{\boldsymbol{\mathbf{y}}}(\theta)$ are measured noisy output and parametric output vectors for $\textbf{N} _{\textbf{exp}}$ experiments. $|\boldsymbol{\Sigma_{\tilde{\boldsymbol{\mathbf{y}}}}(\theta)}|$ is determinant of the covariance matrix. Likelihood function obtained in \eqref{eq:multivariate} as the joint distribution of $\textbf{N} _{\textbf{exp}}$ measurements, can be used in order to update a posterior probability using \eqref{eq:posterior}.
\section{$\stl$ Constraints and Their Under-Approximation}
\label{sec:STL}
\subsection{Signal Temporal Logic ($\stl$)}
\label{subsec:STL}
One of the advantages of $\stl$ specifications is their capabilities in defining temporal specifications for trajectories of physical systems.
We denote an infinite trajectory of the system in \eqref{eq:parametricmodel} by $\xi=x(0)x(1)x(2),\ldots$ where $x(t)$ is the state of the system at time $t\in\mathbb N_0 :=\{0,1,2,\ldots\}$.
\smallskip
\noindent
\textbf{Syntax}: Signal temporal logic ($\stl$) formulae are defined recursively using the following syntax:
\begin{align*}
&\psi ::= \mathsf{T}\mid\mu\mid\neg\psi\mid\psi\wedge\phi\mid\psi\until_{[a,b]}\phi,
\end{align*}
where, $\mathsf{T}$ is the true predicate, and $\mu:\mathbb R^n\rightarrow \{\mathsf T,\mathsf F\}$ is a predicate which its truth value is determined by the sign of a function of the state $x$, i.e., $\mu(x)= \mathsf T$ if and only if $\alpha(x)\ge0$ with $\alpha:\mathbb{R}^n\to\mathbb{R}$ being an affine function of the state and is associated with $\mu$. Notations $\neg$ and $\wedge$ denote negation and conjunction of formulas. Notation $\until_{[a,b]}$ denote the until operator where $a,b\in\mathbb{R}_{\geq{0}}$.
\smallskip
\noindent
 \textbf{Semantics:} The satisfaction of an $\stl$ formula $\psi$ by a trajectory $\xi$ at time $t$ is defined recursively as follows:
\begin{align*}
&(\xi,t)\models\mu\Leftrightarrow\mu(\xi,t)=\mathsf{T}\\
&(\xi,t)\models\neg\mu\Leftrightarrow\neg((\xi,t)\models\mu)\\
&(\xi,t)\models\psi\wedge\phi\Leftrightarrow(\xi,t)\models\psi\wedge(\xi,t)\models\phi\\
&(\xi,t)\models\psi\until_{[a,b]}\phi\Leftrightarrow\exists t'\in[t+a,t+b] ~\text{s.t.}~ (\xi,t')\models\phi\\
&\wedge\forall t''\in[t,t'],~(\xi,t'')\models\psi.
\end{align*}
 A trajectory $\xi$ satisfies a specification $\psi$, denoted by $\xi\models\psi$, if $(\xi,0)\models\psi$.  
 Furthermore, other standard operators can be defined using the above defined operators. For \emph{disjunction}, we can write $\psi\vee\phi:=\neg(\neg\psi\wedge\neg\phi)$ and the \emph{eventually} operator can be defined as $\eventually_{[a,b]}\psi:=\mathsf{T}\until_{[a,b]}\psi$. Finally, the \emph{always} operator is defined as $\always_{[a,b]}\psi:=\neg\eventually_{[a,b]}\neg\psi.$
\medskip 
The \emph{horizon} of an $\stl$ formula denoted by $len(\psi)$ is the maximum over all upper bounds of intervals on the temporal operators. Intuitively, $len(\psi)$ is the horizon in which satisfaction of $(\xi,t)\models\psi$ should be studied.
Let us now denote a finite trajectory by $\xi(t:N):=x(t)x(t+1)...x(t+N)$. For checking $(\xi,t)\models\psi$, it is sufficient to consider a finite trajectory  $\xi(t:N)$ with $N=len(\psi)$. 

\subsection{$\stl$ Robustness}
A real-valued function $\rho^\psi$ can be assigned to an $\stl$ formula $\psi$ such that $\rho^\psi(\xi,t)>0$ implies that $(\xi,t)\models\psi$. This function is called \emph{robustness} of the formula and is defined as follows:
\begin{align*}
\rho^{\mathsf{T}}(\xi,t)&=+\infty,\\
\rho^{\mu}(\xi,t)&=\alpha(\xi(t))\;\;\; \mu(x(t))=\mathsf{T}~\text{iff}~\alpha(x(t))\geq 0,\\
\rho^{\neg\mu}(\xi,t)&=-\rho^{\mu}(\xi,t),\\
\rho^{\psi\wedge\phi}(\xi,t)&=\min(\rho^{\psi}(\xi,t),\rho^{\phi}(\xi,t)),\\
\rho^{\psi\until_{[a,b]}\phi}(\xi,t)&=\max_{i\in[a,b]}(\min(\rho^{\phi}(\xi,t+i), \min_{j\in[0,i)}\rho^{\psi}(\xi,t+j))),
\end{align*}
where $x(t)$ is the value of trajectory $\xi$ at time $t$. As defined above, the robustness of an STL formula can be computed recursively based on its structure and using $\min-\max$ operators.

\begin{remark}
The $\stl$ robustness and satisfaction are defined with respect to a given sample trajectory of the system. When the system is stochastic, the trajectory is a stochastic process, which makes the satisfaction relation a Bernoulli random variable and the robustness a real random variable. In the sequel, we study the probability of satisfying an $\stl$ specification by the stochastic trajectories of the system. 
\end{remark}

\subsection{Under-approximation of $\stl$ Constraints}
The stochastic satisfaction function defined in~\eqref{eq:fsfc} requires the exact feasible set of the chance constraint in \eqref{eq:chance_const}. This feasible set does not have a closed form in general. Previous works tried to find under-approximations of the feasible set. We leverage the proposed procedure in \cite{farahani2018shrinking} to get an under-approximation of the feasible set. This procedure transforms the chance constraints on the $\stl$ property into similar constraints on the predicates of the property using the structure of the $\stl$ formula. We discuss this procedure in this subsection and show how this under-approximation can be improved in Subsection~\ref{subsec:conservativeness}. 

Suppose an $\stl$ formula $\psi$ has a finite horizon $ len(\psi)$.
The robustness of $\psi$ indicates that the trajectory $\xi$ of the system satisfies $\psi$ at time $t$ with probability greater than or equal to $1-\delta$, if $\xi(t:N)=x(t)x(t+1)x(t+2)\cdots x(t+N)$ with $N =  len(\psi)$ satisfies $\textbf{Pr}(\rho^{\psi}(\xi(t:N))>0)\geq 1-\delta$.

The next lemma, borrowed from \cite{farahani2018shrinking}, shows how one can transform the chance constraints on the satisfaction of $\stl$ formulae into similar constraints on the predicates of formulae.
\begin{lemma}\label{lemma:underap}
    For any $\stl$ formula $\psi$ and a value $\delta\in(0,1)$, probability constraints of the forms $\textbf{Pr}(\xi(t:N)\models\psi)\geq{1-\delta}$ and $\textbf{Pr}(\xi(t:N)\models\psi)\leq{1-\delta}$ can be transformed into similar constraints on the  predicates of $\psi$ based on the structure of $\psi$.
\end{lemma}
In the following, we discuss how this transformation is performed.\\
\textbf{ Case I} Negation $\psi=\neg\psi_1$
\begin{align}\label{eq:under11}
\textbf{Pr}(\xi(t:N)\models\neg\psi_1)&\geq\delta \Leftrightarrow\\\notag 
\textbf{Pr}(\xi(t:N)\models\psi_1)&\leq 1-\delta.
\end{align}
\textbf{ Case II} Conjunction $\psi=\psi_1\wedge\psi_2$
\begin{align}
\textbf{Pr}(\xi(t:N)\models\psi_1\wedge\psi_2)&\geq\delta  \Leftarrow\\\notag
\textbf{Pr}(\xi(t:N)\not\models\psi_i)&\leq \frac{1-\delta}{2},~ i=1,2.
\end{align}
\textbf{ Case III} $\psi=\psi_1\until_{[a,b]}\psi_2$\\
\begin{align}\label{eq:underapp4}
\textbf{Pr}(\xi(t:N)\models\psi)&\geq\delta \Leftarrow\\\notag
\textbf{Pr}(\Lambda_j)&\geq\frac{\delta}{(b-a+1)},\,\, j=1,\ldots,N,
\end{align}
in which the events $\Lambda_j$ are defined as 
\begin{align}
\Lambda_j:=\bigwedge\limits _{k=t}^{t+a-1} (\xi(k:N)\models \psi_1)\bigwedge \limits  & _{k=a+t}^{j-1}(\xi(k:N) \models (\psi_1\wedge\neg\psi_2))\nonumber\\&\wedge\xi(j:N)\models\psi_2.
\label{eq:untill}
\end{align}
These transformations are based on multiple applications of Boole's inequality \citep{derler2011modeling}. Required transformations for the complements of Case II and Case III can be derived similarly.


Lemma~\ref{lemma:underap} enables us to write down probabilistic inequalities on the satisfaction of atomic predicates and use them as an under-approximation of the original probabilistic $\stl$ constraint. These probabilistic inequalities can be equivalently written as algebraic inequalities given that we know the statistical properties of the state trajectories. In the case of LTI systems with Gaussian disturbances, $x(t)$ is also Gaussian with known mean and covariance matrix. For the predicate $\mu(x)=\{\alpha(x)\geq0\}$ with $\alpha(x) :=  \tilde{\theta}_0+\tilde{\theta}^T x$, for some $\tilde{\theta}\in\mathbb R^n$ and $\tilde{\theta}_0\in\mathbb R$, we have $\mathbb E[\alpha(x)] = \tilde{\theta}_0+\tilde{\theta}^T \mathbb E[x]$ and $\textrm{Var}[\alpha(x)] = {\tilde{\theta}}^T \textrm{Cov}(x)\tilde{\theta}$. Therefore,
\begin{align}
&\textbf{Pr}(\alpha(x)\geq0)\geq{1-\delta}\; \Leftrightarrow\; 
\textbf{Pr}(\alpha(x)<0)\leq{\delta}\nonumber\\
&\Leftrightarrow \mathbb E(\alpha(x))+\textrm{Var}(\alpha(x))\mathbf{q}^{-1}(\delta)\geq0,
\label{eq:approx1}
\end{align} 
where $\mathbf{q}^{-1}$ is the error inverse function with $\mathbf{q}=\frac{1}{\pi}\int_{-x}^{x}e^{-t^2}$. In the following theorem, we show that the algebraic inequalities of the form \eqref{eq:approx1} are linear with respect to the input.
\addtolength{\textheight}{-3cm}   

\begin{theorem}\label{theo:linear}
Chance constraint $\textbf{Pr}(\alpha(x(t))\geq0)\geq{1-\delta}$, where $\alpha(x) =  \tilde{\theta}_0+\tilde{\theta}^T x$ and $x(t)$ is the trajectory of the stochastic system \eqref{eq:parametricmodel} at time $t$,  
    can be written as the following affine constraint in terms of the input trajectory: 
    \begin{align}
    &\tilde{\theta}_0+\sum_{i=1}^{t}\tilde{\theta}^TA^{i-1}B\;u(t-i+1)
    +\Gamma(\tilde{\theta},\delta)\ge 0,
    \label{eq:maininequality}
    \end{align}
    \label{theo:approximation}
where
\begin{align*}
\Gamma(\tilde{\theta},\delta):=\bigg(\sum_{i=1}^{t}\tilde{\theta}^TA^{i-1}G\;\pmb{\Sigma_{w}}\;G^T(A^T)^{i-1}\tilde{\theta}\bigg)\mathbf{q}^{-1}(\delta),
\end{align*}
and $\pmb{\Sigma_w}$ is the covariance matrix of the process noise.
\end{theorem}

Note that $\Gamma(\tilde{\theta},\delta)$ is a quadratic function of $\tilde{\theta}$ and depends on $\delta$ nonlinearly.
\subsection{A Less Conservative Approximation}
\label{subsec:conservativeness}
The proposed procedure in Lemma~\ref{lemma:underap} for transforming the chance constraints into similar inequalities on atomic predicates can be very conservative. This is due to the fact that constraints of type $\textbf{Pr}(A_1\cup A_2)\le \delta$ is conservatively replaced by inequalities $\textbf{Pr}(A_i)\le \delta/2$, $i=1,2$.
This replacement puts a uniform upper bound on the probability of events $A_i$ and does not create any room for the intersection of these events.
In this subsection, using \emph{intermediate weighting coefficients}, we increase flexibility in the under-approximation and enlarge the feasible set of the probabilistic $\stl$ constraint.
%

This new under-approximation procedure results in new constraints with a larger number of variables.
It is based on the structure of the $\stl$ formula similar to the discussion in the previous subsection and has the following three cases:
\\
\textbf{ Case I}: Disjunction
\begin{align}\nonumber
&\textbf{Pr}(\xi(t:N)\models(\psi_1\vee\cdots\vee\psi_\iota\vee\cdots\vee\psi_{N}))\geq\delta \Leftrightarrow \\\notag
&\textbf{Pr}(\xi(t:N)\models\psi_\iota)\geq\alpha_{\iota}\frac{\delta}{N},\iota\in\{1,\ldots,N\},\\
&~0\leq\alpha_{\iota}\leq1,\;\alpha_1+\cdots+\alpha_{N}=1.\label{eq:improve1}
\end{align}
\textbf{Case II}: Conjunction
\begin{align}\nonumber
&\textbf{Pr}(\xi(t:N)\models(\psi_1\wedge\cdots\wedge\psi_\iota\wedge\cdots\wedge\psi_{N}))\geq\delta \Leftarrow \\\notag
&\textbf{Pr}(\xi(t:N)\not\models\psi_\iota)\leq\beta_{\iota}\frac{1-\delta}{N},\iota\in\{1,\ldots,N\},\\
&~0\leq\beta_{\iota}\leq1,\;\beta_1+\cdots+\beta_{N}=1.\label{eq:improve2}
\end{align}
\textbf{Case III}: Until
\begin{align}\notag
&\textbf{Pr}(\xi(t:N)\models\psi_1\until_{[a,b]}\psi_2)\geq\delta \Leftrightarrow\\\notag
&\textbf{Pr}(\Lambda_j)\geq\gamma_{\iota}\frac{\delta}{(b-a+1)}, \iota\in\mathbb{N},\\
&~0\leq\gamma_{\iota}\leq1,\;\gamma_1+\cdots+\gamma_{N}=1,
\label{eq:improve3}
\end{align}
in which, $\Lambda_j$ is defined as in \eqref{eq:untill}.

In relations \eqref{eq:improve1}-\eqref{eq:improve3}, $\alpha_{\iota}$, $\beta_{\iota}$, and $\gamma_{\iota}$ are intermediate weighting coefficients which regulates the effect of each probabilistic predicate and result in a bigger feasible set. 

One of the advantages of this approach is that if there is some knowledge about the probability of some predicates, it can be leveraged in order to make the under-approximation more precise and accurate. 

 
\section{Verification of probabilistic $\stl$ constraints}
\label{sec:verification}
\subsection{Feasible Set Computation}
After transforming the probabilistic $\stl$ constraints into the algebraic inequalities, as described in Section~\ref{sec:STL},  these inequalities are in the form of \eqref{eq:maininequality} which are linear with respect to the input trajectory and must hold for the whole input range. We use \emph{robust linear programming} to solve those inequalities. Here, the primary robust linear programming problem is converted to another dual linear programming one without a universal quantifier over the target value based on Farkas' lemma \citep{georghiou2019robust}.
   In the next theorem, we show that the feasible set of the probabilistic predicates at each time step can be characterized by a set of constraints at that time step.
\begin{theorem} \label{theo:feasibleset}
Assume that inputs at each time step $t$ are restricted as $\barbelow{l} \leq u(t) \leq \bar{l} ,\;u(t),\barbelow{l},\bar{l}\in \mathbb{R}^m$. The feasible set of each approximated algebraic inequality in \eqref{eq:maininequality} for the whole range of inputs can be characterized by the set of constraints
	\begin{align}\label{eq:main1}
	P^{T}d &\leq b\\\label{eq:main2}
	D^{T}P &= \textbf{f}_{\tilde{\theta}}\\\label{eq:main3}
	P &\geq 0,
	\end{align}
\end{theorem}
where
\begin{equation*}
\resizebox{\linewidth}{!}{$
P^{T}=[P_1,\ldots,P_{2mt}]\in \mathbb {R}^{1 \times 2mt},\;P_k \in \mathbb{R}_{\ge0},\;\forall k\in \{1,\ldots,2mt\},$}
\end{equation*}
\begin{align*}
d=[\bar{l},\barbelow{l},\ldots,\bar{l},\barbelow{l}]^{T}\in \mathbb {R}^{2mt \times 1},
\end{align*}
\begin{equation*}
b=\tilde{\theta}_0+\Gamma(\tilde{\theta},\delta),\;\tilde{\theta}_0\in\mathbb{R},
\end{equation*}
\begin{equation*}
\textbf{f}_{\tilde{\theta}} =\tilde{\theta}^T[A^{t-1}B; A^{t-2}B;\ldots ;B]\in \mathbb{R}^{t\times1},
\end{equation*}
\begin{align*}
\renewcommand{\arraystretch}{.25}
D = \begin{pmatrix}
1 & 0 & \cdots & \cdots & 0 \\
-1     & 0 & \cdots & \cdots & 0 \\
0 & 1 & 0 & \cdots & 0 \\
0     &-1 & 0     & \cdots & 0\\
\vdots & \ddots& \ddots     & \ddots & \vdots\\
0 & 0 & 0    & 0 & 1 \\
0 & 0 & 0    & 0 & -1 \\
\end{pmatrix}\in \mathbb{R}^{2mt \times t}.
\end{align*}
Solving these constraints simultaneously for all predicates of $\stl$ specification in horizon $N$, leads to the feasible set of parameters for the stochastic system $\textbf{S}$ in \eqref{eq:parametricmodel}. The complexity of computation of confidence value in \eqref{eq:confidence} can be tackled using integrating the updated posteriori distribution over this feasible set by virtue of numerical techniques. Two different numerical approaches are described in the next subsection.
\subsection{Confidence Computation Techniques}
\label{subsec:technique}
\textbf{Mont Carlo Method.}
Considering the nonlinearity in the constraints, computation of integral in \eqref{eq:confidence} can be done efficiently using \emph{Monte Carlo techniques}. The idea is to choose $\textbf{N}$ points uniformly from the bounded region of the parameters and using them in the computation of confidence integral in \eqref{eq:confidence} as long as they satisfy all the required constraints in \eqref{eq:main1}-\eqref{eq:main3} for the whole horizon of $\stl$ properties. Now, the confidence integral is a random variable and can be represented as $Q_{\;\textbf{N}}=\frac{V}{\textbf{N}}\sum_{i=1}^{\textbf{N}}K(\tilde{\theta}_i)$, where $K(\tilde{\theta}_i)=f_\psi^{\delta}(\tilde{\theta}_i)\;p(\tilde{\theta}_i\mid\mathcal{D})$ and $V=\int_{\tilde{\theta}}d\tilde{\theta}$. According to Chebyshev's inequality, one has
\begin{equation}
\label{eq:integral}
\textbf{Pr}(|Q_{\;\textbf{N}}-\mathbb{E}[Q_{\;\textbf{N}}]|\leq\varepsilon)\geq1-\frac{\textrm{Var}[Q_{\;\textbf{N}}]}{\varepsilon^{2}},
\end{equation}
for a given $\varepsilon$, in which
$\textrm{Var}[Q_{\;\textbf{N}}]=\frac{V^{2}}{{\textbf{N}}^{2}}\sum_{i=1}^{\textbf{N}}\textrm{Var}[K(\tilde{\theta}_i)]=\frac{V^{2}\delta_{\textbf{K}}^{2}}{\textbf{N}}$ with $\delta_{\textbf{K}}^{2}=\textrm{Var}[K(\tilde{\theta}_i)]$. Finally, we get $\textbf{Pr}(Q_{\;\textbf{N}}-\mathbb{E}[Q_{\;\textbf{N}}]\leq\varepsilon)\geq1-\frac{V^{2}\delta_{\textbf{k}}^{2}}{{\varepsilon^{2}}\textbf{N}}$. By choosing an appropriate $\textbf{N}$ and $\varepsilon$, one can expect an efficient approximation of the confidence integral.

In order to implement the Monte Carlo technique more effectively, one can restrict the search region by solving an optimization problem over the constraints \eqref{eq:main1}-\eqref{eq:main3} in order to find the extreme points for the parameters, therefore, fewer samples are needed to be chosen in this (potentially) smaller region.

\textbf{Confidence Computation Using Piecewise Affine Approximation Of The Nonlinear Constraint.}
Another approach for computing the confidence value in \eqref{eq:confidence} is approximating the nonlinear term $\Gamma(\tilde{\theta},\delta)$ in \eqref{eq:main1} using \emph{piecewise affine} (PWA) functions. Then, linear programming can be used in order to approximate the feasible set. PWA approximations have been used recently in formal approaches in order to deal with the nonlinearity in dynamical systems \citep{bogomolov2015abstraction,sadraddini2018formal}.

Given that $\Gamma(\tilde{\theta},\delta)$ in \eqref{eq:main1} is continuous and twice differentiable, we can partition its domain into polytopic regions, select a nominal value $(\tilde{\theta}_{0_1},\cdots,\tilde{\theta}_{0_p},\delta_0)$ in each region, and rewrite $\Gamma(\tilde{\theta},\delta)$ in each region as:
\begin{align}
&\Gamma(\tilde{\theta},\delta)\in (\tilde{\theta}_1-\tilde{\theta}_{0_1})^{T}\mathscr{M}_1+\cdots+(\tilde{\theta}_p-\tilde{\theta}_{0_p})^{T}\mathscr{M}_p\nonumber\\
&+(\delta-\delta_{0})^{T}\mathscr{N}+\epsilon\mathscr{B},
\label{eq:orig2}
\end{align}
where
$\mathscr{M}_i=\frac{\partial \Gamma(\tilde{\theta},\delta)}{\partial \tilde{\theta}_i} \,\,\text{ and }\,\,
\mathscr{N}=\frac{\partial \Gamma(\tilde{\theta},\delta)}{\partial \delta}$
at nominal point $\tilde{\theta}_{0_1},\cdots,\tilde{\theta}_{0_p},\delta_{0}$
and $\epsilon$ is a bound on
\begin{align*}
&\epsilon\ge \frac{1}{2}[(\tilde{\theta}_1-\tilde{\theta}_{0_1}),\cdots,(\tilde{\theta}_p-\tilde{\theta}_{0_p}),
(\delta-\delta_0)]\;\bm{\mathrm{H}}\\ &[(\tilde{\theta}_1-\tilde{\theta}_{0_1}),\cdots,
(\tilde{\theta}_p-\tilde{\theta}_{0_p}),
(\delta-\delta_0)]^T,
\end{align*}
where $\bm{\mathrm{H}}$ is the hessian matrix and $\mathscr{B}$ denotes the unit interval $[-1,1]$. The region of parameters is divided into sufficiently large number of regions and then inequalities and equations regarding the satisfaction of $\stl$ properties in \eqref{eq:main1}-\eqref{eq:main3} will be checked in these regions. The linearization area can be made smaller than the initial parameter space $\Theta$ by solving an appropriate optimization problem over constraints \eqref{eq:main1}-\eqref{eq:main3} and finding extreme values for the parameters.
In the next lemma, we show that the real feasible set can be constructed in the limit if the number of piecewise regions increases.
\begin{lemma}\label{lemma:pwa}
	The feasible set of \eqref{eq:main1}-\eqref{eq:main3} for all predicates and time steps within the horizon of $\stl$ property (if existing) can be recovered in the limit for large numbers of piecewise regions in order to approximate the nonlinear part of \eqref{eq:main1}.
\end{lemma}

\section{Experimental Results}
\label{sec:casestudy}
Consider a parameterized class of models $\mathrm{M}(\theta)$ with the state-space representation
\begin{equation*}
\resizebox{\linewidth}{!}{$
	\mathrm{M}(\theta)\in \left\{\begin{array}{l}
	x(t+1)=\begin{bmatrix} 
	a && 0\\
	1-a^{2} && a
	\end{bmatrix}x(t)+\begin{bmatrix}
	\sqrt{1-a^{2}}\\
	-a\sqrt{1-a^{2}}
	\end{bmatrix}{u}(t)+\begin{bmatrix} 
	1 && 0\\
	0 && 1
	\end{bmatrix}w(t)\\ 
	~~~\hat{y}(t,\theta)=\theta^{T}x(t).
	\end{array}\right.$}
\end{equation*} 
Each model in $\mathrm{M}(\theta)$ has a single input and a single output. The coefficient $a$ is $0.4$ and the parameter set is selected as $\theta \in \Theta=[-10,10]\times[-10,10]$. The system $\mathbf{S}\in \mathrm{M}(\theta)$ has the true parameter $\theta_{\textit{true}}=[-0.5,1]^T$. System $\mathbf{S}$ is a member of models demonstrated by the Laguerre-basis functions as transfer functions \citep{haesaert2015data}. This is a special case of the orthonormal basis functions and can be translated to the aforementioned parameterized state space format. The system is affected by a process noise which is a Gaussian noise with variances of $0.5$. There is also an additive measurement noise with zero-mean and covariance matrix of $0.5I_2$. The input range is considered to be $[-0.2,0.2]$. 

\begin{figure}[t]
	\centering 
	\includegraphics[width=1\linewidth]{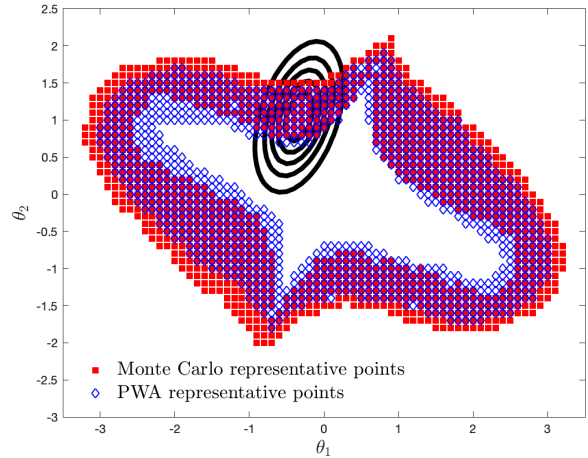}
	\caption{Contours of $p(\theta\mid\mathcal{D})$ for $\theta_{\textit{true}}=[-0.5,1]^T$ after $50$ measurements over the feasible set computed by the Monte Carlo and PWA techniques.}
	\label{fig:ver}
\end{figure}

We want to verify with high probability if the output of the system $\mathbf{S}$ remains in $\mathfrak{l_1}=[-0.5,0.5]$ until it reaches $\mathfrak{l_2}=[-0.1,0.1]$ at some time in the interval $[2,4]$.
We denote the atomic propositions $\mu_1 = \{y\ge-0.5\}$, $\mu_2 = \{-y\ge-0.5\}$, $\mu_3 = \{y\ge-0.1\}$, $\mu_4 = \{-y\ge-0.1\}$.
Our desired property can be written as $\textbf{Pr}(\mathbf{S} \models (\mu_1 \wedge \mu_2)\until_{[2,4]}(\mu_3 \wedge \mu_4))\geq1-\delta$. We select $\delta=0.01$.
The system starts at the initial condition $x(0)=0$, thus the constraint on the initial output already holds. We use the procedure in Section~\ref{sec:STL} to decompose this $\stl$ property to algebraic constraints on the atomic propositions. Equation \eqref{eq:improve2} is used to improve the conservativeness of the approximation.
The feasible set is approximated either using the Monte Carlo method or the piecewise affine approximation described in Section~\ref{sec:verification}. The initial set can be restricted by finding the extreme values of $\theta$ over all constraints as described in Subsection~\ref{subsec:technique} which is considered $[-3.5,3.5]$ for this case study. We select $27947$ points uniformly in this restricted region in order to compute the confidence value using Monte Carlo method with choosing $\varepsilon=0.005$ and $\textrm{Var}[Q_{\;\textbf{N}}]=0.058$ in \eqref{eq:integral}. Computed feasible set using the Monte Carlo technique is demonstrated in Fig.~\ref{fig:ver} with red-face squares. The feasible set which is recovered with the piecewise affine technique is illustrated in Fig.~\ref{fig:ver} with blue-edge diamonds. We used linear programming in order to find the feasible set of parameters for the linearized form of \eqref{eq:main1} together with \eqref{eq:main2} and \eqref{eq:main3} for all time steps in $\theta$ and $P$ space. Then, this feasible set is projected into $\theta$ space using MPT3 toolbox \citep{herceg2013multi}. We choose the total number of regions in the piecewise affine approximation to be $25$.

As we do not have any prior knowledge about the parameters, we choose a uniform distribution $p(\theta)$ on the possible models. Based on the uniform prior, the confidence is computed using \eqref{eq:confidence} as $0.0279$ and $0.0258$ with Monte Carlo and PWA approximations, respectively. Afterward, we designed an experiment on the system with the true parameter and an input sequence as Gaussian noise with a uniform distribution over $[-2,2]$ and measured output for $50$ consecutive time instances. Using updated $p(\theta\mid\mathcal{D})$ coming from the measurement data, confidence improved significantly into $0.9099$ and $0.8962$ for Monte Carlo and PWA, respectively. We repeated the same experiment $100$ times for several other true parameters $\theta_{\textit{true}}$. For all of these instances, updated posteriori probability in \eqref{eq:multivariate}, after $50$ measurements, is used in order to compute the confidence value according to \eqref{eq:confidence}. Contours of the posterior distribution are illustrated in Fig.~ \ref{fig:ver} . Results of computing the confidence with Monte Carlo and PWA approximation are shown in Table~\ref{tab:table1}. As it can be seen, for parameters that lie deep inside the feasible set, the confidence value is high with a low variance for both techniques. Meanwhile, for the points near the edges, the variance is higher and confidence value is lower. For points far enough from the feasible set, confidence tends to be zero.
\\
\vspace*{-1.5mm}
\begin{table}[h]
	\renewcommand{\arraystretch}{1}
	\text{\textbf{Table1.} Means and variances of computed confidence} \\ \text{ values for $5$ different true parameters.}
	\begin{center}
		\begin{tabular}{ccccc}
			\arrayrulecolor{pink}
			\toprule
			\multicolumn{3}{r}{Monte Carlo\;\;\;\;\;\;} &          \multicolumn{2}{c}{PWA\;\;} \\
			\cmidrule(r){2-3}
			\cmidrule(r){4-5}
			\cellcolor{white}$\theta_{\textit{true}}$
			& \cellcolor{white} Mean & \cellcolor{white}Variance & \cellcolor{white}Mean & \cellcolor{white}Variance \\
			\midrule
			$[-0.5,1]^T$     &       0.9587   & 0.0023 & 0.9514 & 0.0042     \\
			$[3,-1]^T$&   0.4902    & 0.0061  & 0.5032 & 0.0062   \\
			$[1,0.5]^T$     & 0.7932    &0.0025 & 0.7584 & 0.0053    \\
			$[-2,1.5]^T$      & 0.9018   & 0.0009  & 0.9156 & 0.0005  \\
			$[2,-1]^T$ &0.0278   & 0.0005   & 0.0480 & 0.0006  \\
			\bottomrule
		\end{tabular}
	\end{center}
	\label{tab:table1}
\end{table}

\section{Conclusion and Future Works}
In this work, we considered stochastic dynamical systems that do not have a precise model but a parametric model is available. We developed a scheme for computing confidence value for satisfaction of properties expressed in signal temporal logic formuale
using both model-based methods and Bayesian inference techniques.
Our approach transforms the temporal property into algebraic inequalities.
By leveraging the collected data from the system, the probability density of the unknown parameters is updated and the confidence value is computed over the feasible domain of the parameters.
Two numerical techniques, Monte Carlo and piecewise affine approximation, are used for the computation.
 Future work will be concentrated on investigating the implementation of the maximum likelihood concept in order to excite the system with inputs that maximize the probability of observations.  Furthermore, we are also interested in combining the Bayesian data analysis and model-based techniques in real time.

\bibliography{sample} 


                                                   






%
\end{document}